\documentclass[journal]{IEEEtran}
\pdfoutput=1 
\usepackage{cite}
\usepackage{subfigure}

\usepackage{amsmath,amssymb,amsfonts}
\usepackage{graphicx}
\usepackage{overpic}

\newcommand{\bitm}{\begin{itemize}}
\newcommand{\eitm}{\end{itemize}}
\newcommand{\beqa}{\begin{eqnarray}}
\newcommand{\eeqa}{\end{eqnarray}}
\newcommand{\beqas}{\begin{eqnarray*}}
\newcommand{\eeqas}{\end{eqnarray*}}
\newcommand{\baln}{\begin{align}}
\newcommand{\ealn}{\end{align}}
\newcommand{\balns}{\begin{align*}}
\newcommand{\ealns}{\end{align*}}

\newcommand{\probSimplex}[1] {\mathcal{P}\parenth{#1}}

\newcommand{\kldist}[2] {D\! \parenth{#1\|#2}}
\newcommand{\E}{\mathbb{E}}
\renewcommand*{\P} {\mathbb{P}}

\newcommand{\Q} {\mathbb{Q}}

\newcommand{\cF}{\mathcal{F}}
\newcommand{\cB}{\mathcal{B}}

\newcommand{\alphabet}[1] { {\mathsf #1}}

\newcommand{\R}{\mathbb{R}}

\newcommand{\reals}{\R}

\newcommand{\parenth}[1] {\left(#1\right)}

\newcommand{\brackets}[1] {\left[#1\right]}

\newcommand{\pmf}[2] {P_{#1}\!\parenth{#2}}

\newcommand{\I}{\text{I}}

\newcommand{\D}{\text{D}}

\newcommand{\X}{\bX}
\newcommand{\Y}{\mathbf{Y}}
\newcommand{\Z}{\mathbf{Z}}
\newcommand{\N}{\mathbf{N}}
\newcommand{\bfI}{\mathbf{I}}

\def\argmin{\mathop{\arg\,\!\min}\limits}%
\def\argmax{\mathop{\arg\,\!\max}\limits}%

\newcommand{\calX}{\alphabet{X}}

\newcommand{\calZ}{\alphabet{Z}}
\newcommand{\bX}{\mathbf{X}}
\newcommand{\bY}{\mathbf{Y}}
\newcommand{\bZ}{\mathbf{Z}}
\newcommand{\bx}{\mathbf{x}}
\newcommand{\by}{\mathbf{y}}
\newcommand{\bz}{\mathbf{z}}
\newcommand{\allX}{\underline{\bX}}
\newcommand{\allx}{\underline{\bx}}

\newcommand{\PT}{\widehat{P}_{\allX}}

\newcommand{\A}{\textbf{A}}

\newcommand{\mT}{\mathcal{T}}
\newcommand{\mcT}{\mathcal{T}_{\text{C}}}

\newcommand{\tildeP}{\widetilde{P}_{\allX}}

\begin{document}
\title{Causal Dependence Tree Approximations of Joint Distributions for Multiple Random Processes}

\author{Christopher~J.~Quinn,~\IEEEmembership{Student Member,~IEEE,} Todd~P.~Coleman,~\IEEEmembership{Member,~IEEE,} and~Negar~Kiyavash,~\IEEEmembership{Member,~IEEE}%
\thanks{The material in this paper was presented (in part) at the International Symposium on Information Theory and Applications, Taichung, Taiwan, October 2010.}
\thanks{C. Quinn is with the Department of Electrical and Computer Engineering, University of Illinois at Urbana Champaign, Urbana, Illinois 61801 (email: quinn7@illinois.edu).  He was supported by the NSF IGERT fellowship, and the Department of Energy Computational Science Graduate Fellowship, which is provided under grant number DE-FG02-97ER25308.}
\thanks{T. Coleman is with the Department of Electrical and Computer Engineering, University of Illinois at Urbana Champaign, Urbana, Illinois 61801 (email: colemant@illinois.edu). }
\thanks{N. Kiyavash is with the Department of Industrial and Enterprise Systems Engineering, University of Illinois at Urbana Champaign, Urbana, Illinois 61801 (email: kiyavash@illinois.edu). }
}


\maketitle

\begin{abstract}

We investigate approximating joint distributions of random processes with causal dependence tree distributions. Such distributions are particularly useful in providing parsimonious representation when there exists causal dynamics among processes. By extending the results by Chow and Liu on dependence tree approximations, we show that the best causal dependence tree approximation is the one which maximizes the sum of directed informations on its edges, where best is defined in terms of minimizing the KL-divergence between the original and the approximate distribution. Moreover, we describe a low-complexity algorithm to efficiently pick this approximate distribution.

\end{abstract}

\IEEEpeerreviewmaketitle

\section{Introduction}

For many problems in statistical learning, inference, and prediction, it is desirable to find a parsimonious representation of the full joint distribution of multiple random processes with various interdependencies.  Such an approximation of the joint distribution can lend itself both to easier analysis and inference, as well as reduced storage requirements. More importantly, parsimonious representations facilitate visualization and human comprehension of data. Specifically, in situations such as network intrusion detection, decision making in adversarial environments, and first response tasks where a rapid decision is required, such representations can greatly aid the situation awareness and decision making process. 


To facilitate analysis and visualization, graphical representations are used to describe both the full and the approximating distributions \cite{pearl2009causality, lam1994learning, friedman1998learning, friedman2003being, heckerman1996bayesian, koivisto2004exact, cheng2002learning, murphy2002dynamic, pearl1988probabilistic, heckerman2008tutorial}. In such representations, variables are represented as nodes and undirected edges between each pairs of variables depict statistical dependence. Therefore,  a variable is statistically independent of all of the variables it does not share an edge with  \cite{heckerman2008tutorial}.  

One of the simplest graph structures is a tree.  A tree is a connected graph on $n$ nodes which has $n-1$ edges, and consequently has no loops.  Dependence tree approximations are comparatively simple to analyze (few dependencies retained) and require significantly less storage requirements (storing the full joint requires exponential space in the number of variables; dependence trees  require linear space).  

There are many choices for tree approximations, and often a criterion, such as Kullback-Leibler (KL) divergence, is used to define ``goodness.''  Chow and Liu showed that dependence tree approximation with the minimum KL divergence was the one that maximized the sum of the mutual informations between variables sharing an edge  \cite{chow1968approximating}.  They also identified a low complexity algorithm, based on minimum spanning tree algorithm, to identify this best tree \cite{chow1968approximating}.  Their proposed algorithm only  requires the computation of second order distributions (pairwise interactions) find the best approximation of the whole joint density.

\begin{figure}[t]
\centering
\includegraphics[width=.8\columnwidth]{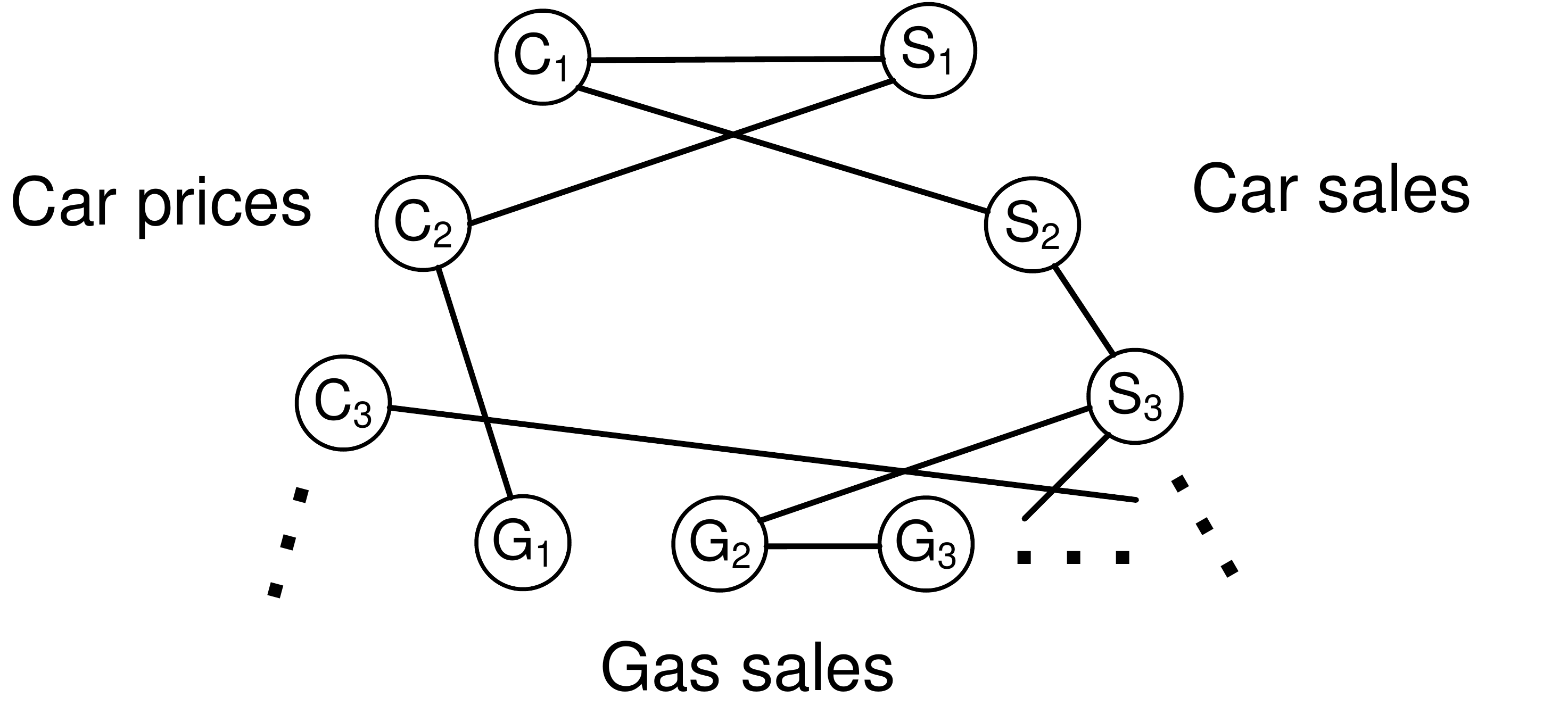}
\caption{A possible result in applying Chow and Liu's work to the example of car prices, car sales, and gas sales ($C^{365}$, $S^{365}$, and $G^{365}$ respectively) over a year.  With over one thousand variables, the structure gets increasingly complex with time. Most importantly, even though the system dynamics are causal, the tree approximation is not.}
\label{fig:intro:chow_liu_graph}
\end{figure}

For some learning and inference problems, it might be desirable to have models which keep the temporal structure.  Directly applying Chow and Liu's procedure to multiple random processes can yield approximations which do not preserve temporal structure and which become increasingly complex with time.  This can be demonstrated with an example.  Consider the problem of identifying a simple but meaningful summary of how car prices $\{C_1, C_2, \dots, C_{365}\}$, the number of cars sold $\{S_1, S_2, \dots, S_{365}\}$, and gas sales $\{G_1, G_2, \dots, G_{365}\}$ in a town change over the course of a year.  Suppose we have access to the full joint distribution $P_{C^{365}, S^{365}, G^{365}}(c^{365}, s^{365}, g^{365})$.  One possible result is shown in Figure~\ref{fig:intro:chow_liu_graph}.  This figure only shows the beginning of the processes; there are over one thousand nodes in this tree.  Even though this graph does not have many edges for the number of nodes present (much simpler than the full joint), it has a complicated structure, making analysis difficult.  With increasing time, it would become more complicated.

Also, this approximation, like almost all other possible Chow and Liu approximations for this problem, does not preserve temporal ordering.  If we tried to interpret causal dependencies between the variables shown in Figure~\ref{fig:intro:chow_liu_graph} as is done in \cite{pearl2009causality, lam1994learning, friedman1998learning, friedman2003being, heckerman1996bayesian, koivisto2004exact, cheng2002learning, murphy2002dynamic, pearl1988probabilistic, heckerman2008tutorial}, we would conclude that either a) the price of cars on day one ($C_1$) depends on car sales on day two ($S_2$) and gas on day one ($G_1$) depends on the price of cars on day two ($C_2$) or b) car sales on day one ($S_1$) depends on the number of cars sold on day two ($C_2$).  In either case, a process on day one is seen to depend on another process on day two.  For real world examples with  \emph{causal} dynamics, the present might depend on the past, but not the future.  While this approximation can be easily used to infer correlative influences, it might be difficult to infer causal influences  from it.

\begin{figure}[t]
\centering
\includegraphics[width=.8\columnwidth]{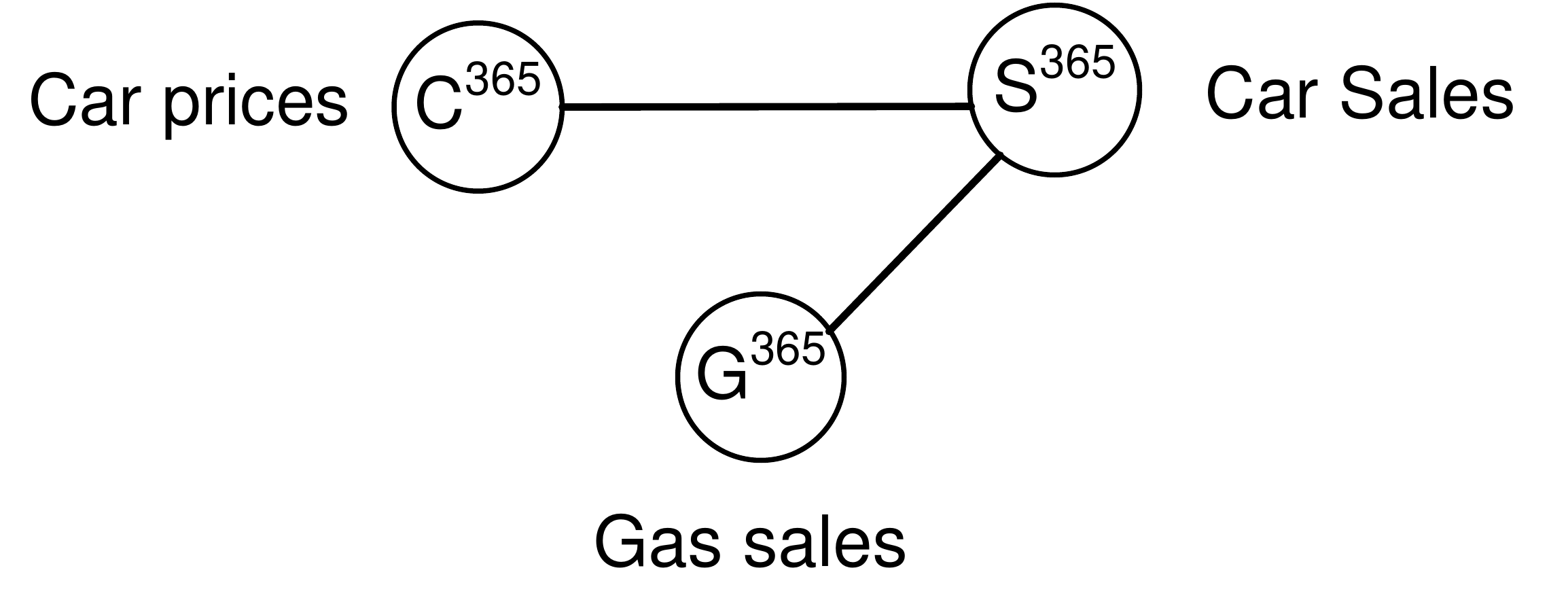}
\caption{A possible result of applying Chow and Liu's work to the example of car prices, car sales, and gas sales ($C^{365}$, $S^{365}$, and $G^{365}$ respectively) over a year, where each process is treated as a random object.  The graphical complexity is low and does not grow with time.  However, no causal relationships can be inferred, only correlative ones.}
\label{fig:intro:ex_simple_undir}
\end{figure}

Although directly applying the Chow and Liu procedure to multiple random processes might result in an approximation with undesirable properties, there is an alternative way to apply the procedure.  Consider treating each \emph{process} as a random object.  
A possible Chow and Liu approximation of this for the example above is shown in Figure~\ref{fig:intro:ex_simple_undir}.  With this technique, the complexity is low for all time and the processes are kept intact.  Consequently, inferring  relationships between the processes is much simpler.  However, since all of the time steps are kept together, still no \emph{causal} influences can be inferred and only correlative relationships can be recovered.

\begin{figure}[t]
\centering
\includegraphics[width=.8\columnwidth]{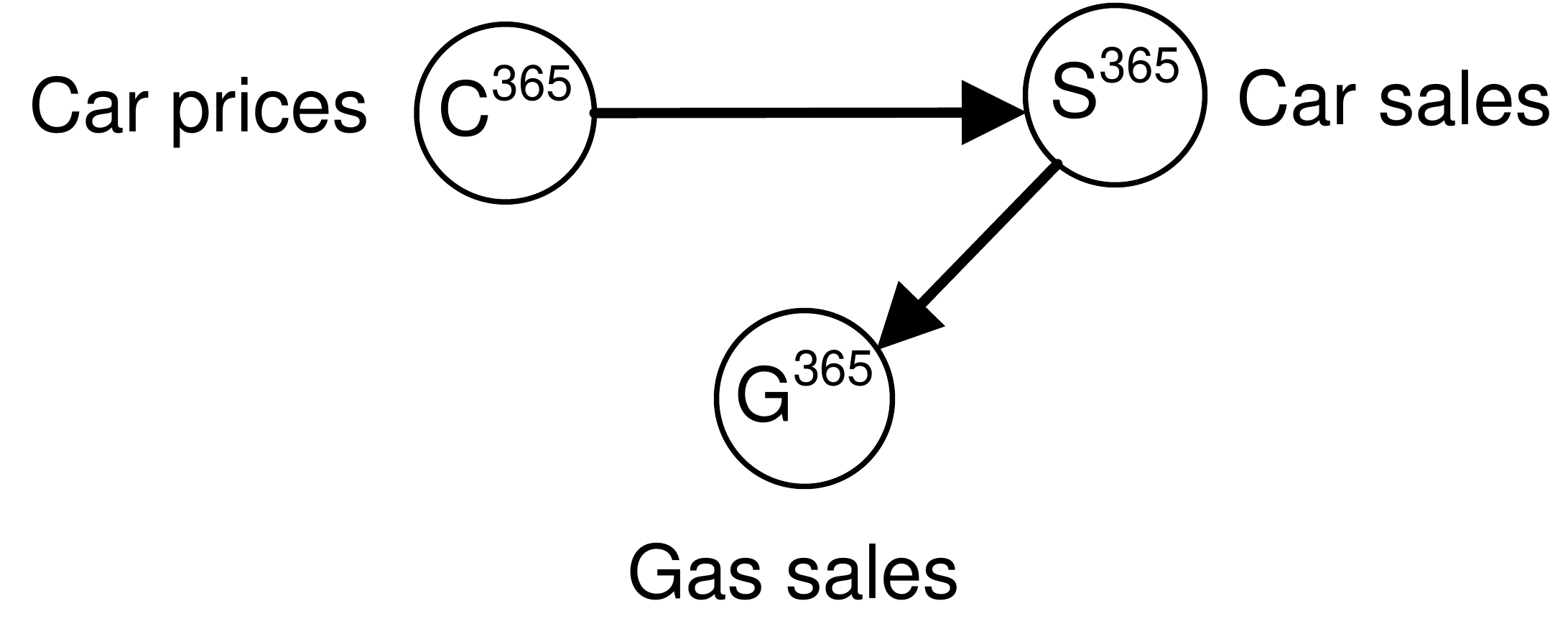}
\caption{A possible causal dependence tree approximation for the example of car prices, car sales, and gas sales ($C^{365}$, $S^{365}$, and $G^{365}$ respectively) over a year.  The graphical complexity is low and does not grow with time.  The dependence tree is \emph{causal}, which is important since the underlying system being approximated is also causal.}
\label{fig:intro:ex_simple_dir}
\end{figure}

\section{Our contribution and related work}

\subsection{Our Contribution}

In this paper, we develop a procedure similar to Chow and Liu's, but in the context of random processes.  Our approach is motivated by approximating real world dynamical systems, where there are physical, causal relationships.  Our approach recovers a parsimonious causal tree representation that approximates the original system dynamics. The goodness of the approximation is measured by KL divergence.  We show that the causal dependence tree approximation with the minimum KL divergence is the one that maximizes the sum of the pairwise directed informations between processes sharing an edge. This allows us to present a low complexity maximum weight directed spanning tree algorithm for calculating the best approximate causal tree.

Such a tree, as demonstrated in Figure~\ref{fig:intro:ex_simple_dir} for the example regarding car prices, car sales, and gas sales,  can be represented graphically with directed edges corresponding to the direction of influence.  Besides maintaining the causal dynamics, which is a property of most real systems, our proposed approach does not suffer from quick growth of complexity with time, as do \cite{pearl2009causality, lam1994learning, friedman1998learning, friedman2003being, heckerman1996bayesian, koivisto2004exact, cheng2002learning, murphy2002dynamic, pearl1988probabilistic, heckerman2008tutorial} (Figure~\ref{fig:intro:chow_liu_graph}), since it works with random processes which are not intermixed, like in Figure~\ref{fig:intro:ex_simple_undir}.

\subsection{Related work}

There is a large body of work on approximating joint distributions with probabilistic graphical models, which are often called Bayesian networks \cite{pearl2009causality, lam1994learning, friedman1998learning, friedman2003being, heckerman1996bayesian, koivisto2004exact, cheng2002learning, murphy2002dynamic, pearl1988probabilistic, heckerman2008tutorial}.  Chow and Liu were the first researchers in this field to investigate \emph{tree} approximations \cite{chow1968approximating} for discrete random variables.  
Suzuki extended the result to general random variables \cite{suzuki2010generalization}.  Carvalho and Oliveira considered Chow and Liu's problem for metrics other than KL divergence \cite{carvalho2007learning}.   Meila and Jordan generalized the Chow and Liu procedure to find the best mixture-of-trees approximation \cite{meila2001learning}.  Choi et al. developed methods based on Chow and Liu's to learn dependence tree approximations of distributions with hidden variables \cite{choi2010learning}.



The work in Bayesian networks largely addresses correlative relationships, not causal ones.  There has been work in developing methods to identify statistically causal relationships between processes.  When the processes can be modeled by multivariate auto-regressive models, Yuan and Lin developed a method, ``Group Lasso,'' which can be used to infer the causal relationships \cite{yuan2006model}.  Bolstad et al.  recently showed conditions under which the estimates of Group Lasso are consistent and propose modifications to improve the reliability \cite{nowak2010causal}.  Materassi has developed methods based on Wiener filtering to infer statistically causal influences in linear dynamic systems.  Consistency results have been derived for the case when the underlying dynamics have a tree structure \cite{materassi2010topological, materassi2010problem}.

Granger proposed a widely adopted framework for identifying causal influences based on statistical prediction \cite{granger1969investigating}.
There have been a number of proposed quantitative measures based on this.  There are many based on Granger's original measure based on linear models \cite{granger1969investigating}, but will not be referenced here.  In the context of dynamical systems, Marinazzo et al. developed a measure of Granger causality based on kernel methods for multiple processes \cite{marinazzo2008kernel}. Massey and Rissanen independently proposed a measure, \emph{directed information} \cite{rissanen1987measures,massey1990causality }, which is based on earlier work by Marko \cite{marko1973bidirectional}.  Solo presented an alternative measure of statistical causality similar to directed information which uses analysis of deviance \cite{solo2009causality}.

There have been some applications of directed information.  Quinn et al. used directed information estimates to infer causal relationships between between simultaneously recorded neurons \cite{quinn2010estimating}.  Rao et al. used directed information estimates to infer causal relationships in gene regulatory networks \cite{rao2007motif}.  In addition to its use in identifying statistically causal influences, directed information also plays a fundamental role in communication with feedback \cite{marko1973bidirectional, kramer1998directed, tatikonda2009capacity, permuter2009finite, massey1990causality,massey2005conservation}, prediction with causal side information \cite{rissanen1987measures, quinn2010estimating}, gambling with causal side information \cite{permuter2008directed, permuter2009interpretations}, control over noisy channels \cite{ elia2004when, martins2008feedback, tatikonda2000control, gorantla2010reversible, tatikonda2009capacity}, and source coding with feed forward \cite{venkataramanan2007source, permuter2009interpretations}.

\subsection{Paper organization}
The paper organization is as follows. In Section III, we establish definitions and notations.  In Section IV, we discuss the problem setup of developing meaningful approximations for a joint distribution of random variables and review the result of Chow and Liu \cite{chow1968approximating}.  In Section V,  we discuss approximating dynamical systems to motivate our approach to solving the problem.  In Section VI, we present our main result of finding the causal dependence tree approximation which best approximates the full joint with respect to KL divergence.  In Section VII, we discuss a low complexity algorithm to identify this best causal dependence tree approximation.  In Section VIII, we analyze properties of causal dependence trees, such as the number of variable dependencies kept and storage requirements, as compared to the full joint distribution and Chow and Liu dependence tree approximations.  In Section IX, we evaluate the performance of causal dependence tree approximations in a binary hypothesis test example, in comparison with the full distributions and Chow and Liu dependence tree approximations.

\section{Definitions and Notation}
This section presents probabilistic notations and information-theoretic definitions and identities that will be used throughout the remainder of the manuscript.  Unless otherwise noted, the definitions and identities come from Cover \& Thomas \cite{cover2006elements}.
\begin{itemize}
\item For a sequence $a_1,a_2,\ldots$, denote $a_i^j$ as $(a_i,\ldots,a_j)$ and $a^k \triangleq a_1^k$.
\item Denote the set of permutations $\pi$ on $\{1,\ldots,m\}$ as $\Pi(m)$.
\item For any Borel space $\calZ$, denote its Borel sets by $\cB(\calZ)$ and the space of probability measures on 
$\parenth{\calZ,\cB(\calZ)}$ as $\probSimplex{\calZ}$.
\item Consider two probability measures $\P$ and $\Q$ on $\probSimplex{\calZ}$.  $\P$ is absolutely continuous with respect to $\Q$ ($\P \ll \Q$) if
$\Q(A) = 0$ implies that $\P(A) = 0$ for all $A \in \cB(\calZ)$.
If $\P \ll \Q$, denote the Radon-Nikodym derivative as the random variable $\frac{d\P}{d\Q}: \calZ \to \reals$ that satisfies
\[ \P(A) = \int_{z \in A} \frac{d\P}{d\Q}(z) \Q(dz),\;\;\; A \in \cB(\calZ).\]
\item 
The {\it Kullback-Leibler divergence} between $\P \in \probSimplex{\calZ}$ and $\Q \in \probSimplex{\calZ}$ is defined as
\beqa
\kldist{\P}{\Q} \triangleq \E_\P \brackets{\log \frac{d\P}{d\Q}} = \int_{z \in \calZ} \log \frac{d\P}{d\Q}(z) \P(dz) \label{eqn:defn:divergence}
\eeqa
if $\P \ll \Q$ and $\infty$ otherwise.
\item Throughout this paper, we will consider $m$ random processes where the $i$th (with $i \in \{1,\ldots,m\}$) random process at time $j$ (with $j \in \{1,\ldots,n\}$), takes values in a Borel space $\calX$.  
\item For a sample space $\Omega$, sigma-algebra $\cF$, and probability measure $\P$, 
      denote the probability space as $\parenth{\Omega,\cF,\P}$.
\item Denote the $i$th random variable at time $j$ by $X_{i,j}: \Omega \to \calX$, the $i$th random process as $\bX_i = (X_{i,1},\ldots,X_{i,n}): \Omega \to \calX^n$, and the whole collection of all $m$ random processes as $\allX= (\bX_1,\ldots,\bX_m)^T: \Omega \to \calX^{mn}$.  
\item  The probability measure $\P$ thus induces a probability distribution on $X_{i,j}$ given by $\pmf{X_{i,j}}{\cdot} \in \probSimplex{\calX}$,
a joint distribution on $\bX_i$ given by $\pmf{\bX_{i}}{\cdot} \in \probSimplex{\calX^n}$, and a joint distribution on $\allX$ given by
$\pmf{\allX}{\cdot} \in \probSimplex{\calX^{mn}}$.  
\item With slight abuse of notation, denote $\bX \equiv \bX_i$ for some $i$ and $\bY \equiv \bX_j$ for some $i \neq j$ and denote the conditional distribution and {\it causally conditioned} distribution of $\bY$ given $\bX$ as
\beqa
P_{\bY|\bX=\bx}(d\by) &\triangleq& \pmf{\bY|\bX}{d\by|\bx} \nonumber \\
 &=& \prod_{i=1}^n \pmf{Y_i|Y^{i-1},X^n}{dy_i|y^{i-1},x^n} \label{eq:def:cond_distr}\\
P_{\bY\|\bX=\bx}(d\by) &\triangleq& \pmf{\bY\|\bX}{d\by\|\bx} \nonumber \\
&\triangleq& \prod_{i=1}^n \pmf{Y_i|Y^{i-1},X^{i}}{dy_i|y^{i-1},x^{i}} \label{eq:def:causal_cond}.
\eeqa
Note the similarity with regular conditioning in \eqref{eq:def:causal_cond}, except in causal conditioning the future ($x^n_{i+1}$) is not conditioned on \cite{kramer1998directed}.
 The notation for $P_{\bY|\bX=\bx}$ and $P_{\bY\|\bX=\bx}$ is used to emphasize that 
 $P_{\bY|\bX=\bx} \in \probSimplex{\calX^n}$ and $P_{\bY\|\bX=\bx} \in \probSimplex{\calX^n}$.
\item
The mutual information and {\it directed information} \cite{massey1990causality} between random process $\bX$ and random process $\bY$ are given by
\beqa
I(\bX;\bY) &=& \int_{\bx} \kldist{P_{\bY|\bX=\bx}}{P_{\bY}} P_{\bX}(d\bx) \label{eqn:defn:MutualInformation}
\\
I(\bX \to \bY) &=& \int_{\bx} \kldist{P_{\bY\|\bX=\bx}}{P_{\bY}} P_{\bX}(d\bx) \label{eqn:defn:DirectedInformation}
\eeqa
Conceptually, mutual information and directed information are related.  However, while mutual information quantifies statistical correlation (in the colloquial sense of statistical interdependence), directed information quantifies statistical \emph{causation}.  For example,
$I(\bX;\bY)=I(\bY;\bX)$, but $I(\bX \to \bY) \neq I(\bY \to \bX)$ in general.
\end{itemize}

\section{Background: Chow and Liu Dependence Tree Approximations}
\begin{figure}[t]
\centering
\includegraphics[width=.6\columnwidth]{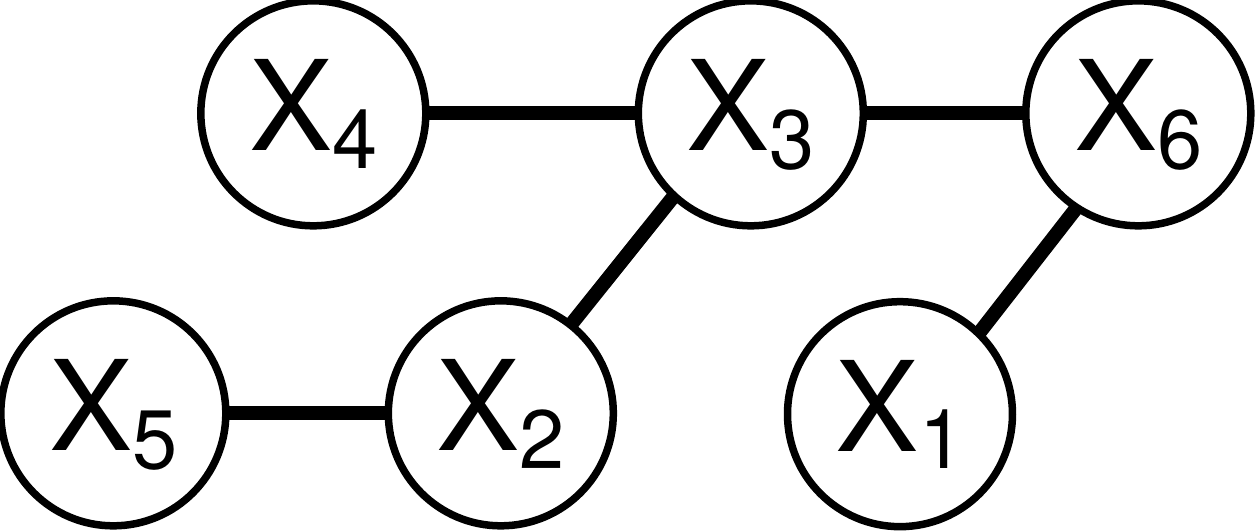}
\caption{Diagram of an approximating dependence tree structure.  In this example,
$\widehat{P}_{X^6}(dx^6) = P_{X_6}(dx_6) P_{X_1|X_6}(dx_1|x_6) P_{X_3|X_6}(dx_3|x_6) P_{X_4|X_3}(dx_4|x_3)$ $\times P_{X_2|X_3}(dx_2|x_3) P_{X_5|X_2}(dx_5|x_2)$. }
\label{fig:dependence_tree}
\end{figure}

Consider the scenario where there are $m$ random processes and there is no time axis (e.g. $n=1$).  Then this becomes a set of just 
$m$ random variables $X^m = \{ X_1, X_2, \cdots, X_m \}$ on $\calX^m$.  Note that the chain rule is given by
 \begin{align}
\!&\!\!\!\pmf{X^m }{dx^m} =\prod_{i=1}^m \!\!\pmf{X_i|X^{i-1}}{dx_i|x^{i-1}} \nonumber \\
\!\!\!&\!\!\!\!\!\quad= \prod_{i=1}^m \!\!\pmf{X_{\pi(i)}|X_{\pi(i-1)},\ldots,X_{\pi(1)}}{ dx_{\pi(i)}| x_{\pi(i-1)},\!\ldots\!,x_{\pi(1)} } \!, \label{eqn:productdistribution:permutation}
\end{align}
where $X^0 \triangleq \emptyset$ and \eqref{eqn:productdistribution:permutation} holds for any permutation $\pi \in \Pi(m)$.
Chow and Liu developed an algorithm to approximate a full joint distribution by a product of second order distributions \cite{chow1968approximating}.  For their procedure, the chain rule is applied to the joint distribution, and each individual term in the product
\eqref{eqn:productdistribution:permutation} is approximated as $\pmf{X_{\pi(i)}|X_{\pi(l(i))}}{dx_{\pi(i)}|x_{\pi(l(i))}}$ where $l(i) \in \{1, \cdots, i-1 \}$, such that the conditioning is on at most one variable.  
This approximation corresponds to a dependence tree structure (see Figure~\ref{fig:dependence_tree}).  Each choice of $\pi(i)$ and $l(i)$ over $\{1, \cdots, m\}$ completely specifies a tree structure $T$.  Denote the set of all possible trees by $\mT$ and the tree approximation of $\pmf{X^m}{x^m}$ using $T \in \mT$ by $ \widehat{P}_{X^m}(dx^m)$:
\begin{eqnarray}
\widehat{P}_{X^m}(dx^m) \triangleq \prod_{i=1}^m \pmf{X_{\pi(i)}|X_{\pi(l(i))}} {dx_{\pi(i)}|x_{\pi(l(i))}}.
\end{eqnarray}

Chow and Liu's method obtains the ``best'' tree $T \in \mT$, where the ``goodness'' is defined in terms of KL distance between the original distribution and the approximating distribution. They show the important property \cite{chow1968approximating}: \\ \newtheorem{chow_liu}{Theorem}
\begin{chow_liu}
\label{chow_liu}
\begin{eqnarray}
\!\!\!\!\!\!\! \argmin_{T \in \mT}   \D ( P_{X^m} \parallel \widehat{P}_{X^m} ) \!\!\!\!\! &=& \!\!\!\!\! \argmax_{T \in \mT}  \sum_{i=1}^m \I(X_{\pi(i)};X_{\pi(l(i))}\!). \label{eq:chow_liu}
\end{eqnarray}
\end{chow_liu}
See \cite{chow1968approximating} for the original proof for discrete random variables, and \cite{suzuki2010generalization} for a proof for general random variables.  The optimization objective is equivalent to maximizing a sum of mutual informations.  Thus, a global minimization is equivalent to (coupled) local maximizations.

They also propose an efficient algorithm to identify this approximating tree by calculating the mutual information between each pair of random variables and assigning those values as weights in the corresponding dependency graph \cite{chow1968approximating}.  Finding the dependence tree distribution that maximizes the sum \eqref{eq:chow_liu} is equivalent to finding a tree of maximal weight in the underlying weighted graph \cite{chow1968approximating}.  Kruskal's minimum spanning tree algorithm \cite{kruskal1956shortest} can be used for this \cite{chow1968approximating}.  The total runtime of this procedure is $\mathcal{O}(m^2)$, where $m$ is the number of random variables (vertices in the graph).

A significant aspect of this result is that only the pairwise interactions need to be known or estimated in order to find the best approximation for the full joint.  In many cases, the statistics of the data are initially unknown.  Chow and Liu's procedure is particularly beneficial when the number of variables is large and, consequently, estimating the full joint distribution is prohibitive.  A simple estimation scheme using empirical frequencies of i.i.d. data is described in \cite{chow1968approximating}.

In \cite{chow2002consistency}, the authors show that if the joint distribution has a dependence tree structure, and if a sufficiently large number of i.i.d. samples are used, then with probability one the estimated tree will be the true joint. Recently, researchers have performed an error exponent analysis for estimating joint distributions with dependence tree structures.  They showed that the error exponent of the probability of the estimated tree structure differing from the true tree structure is equal to the exponential rate of decay of a single dominant ``crossover'' event \cite{tan2009large,tan2009large_journ}.  This event occurs when a pair of non-neighbor nodes in the true tree structure share an edge in the estimated tree structure.

\section{Motivating Example: Approximating the Structure of Dynamical Systems}

As discussed in the introduction, there are potential problems with Chow and Liu dependence tree approximations - the processes could be intermixed and temporal structure might not be kept, as well as an increasing complexity with time.  We now consider how to not only keep the processes unmixed and complexity low, but also to identify causal dependencies between the processes.  To gain intuition for how to approach this problem, we consider the structurally analogous problem of approximating real-world dynamical systems, which evolve through time.

\begin{figure}[t]
\centering
\mbox{\subfigure[Full causal dependence structure]{\label{fig:full_xyz}\includegraphics[width=.35\columnwidth]{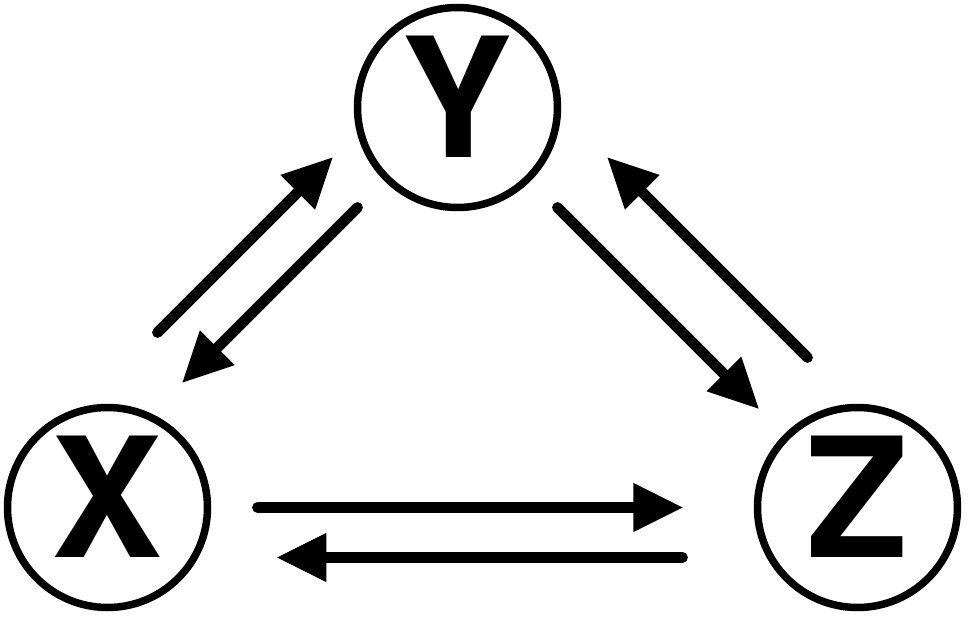}}\qquad\qquad
\subfigure[Causal dependence tree approximation]{\label{fig:tree_xyz}\includegraphics[width=.35\columnwidth]{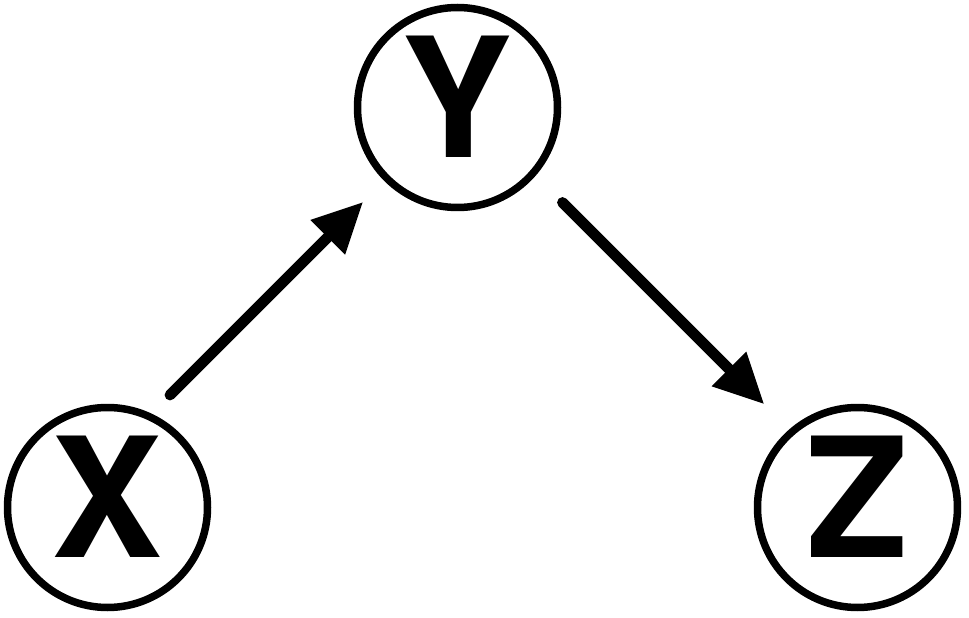}}}
\label{fig:dyn_sys_xyz}
\caption{Dependence tree structures for the dynamical system.}
\end{figure}

Consider approximating a physical, dynamical system.  Such a system evolves causally with time according to a set of coupled differential equations.   Specifically, consider a system with three processes, $\{x_t, y_t, z_t\}$, which evolve according to:
\begin{eqnarray*}
x_{t+\Delta} \!\!\!\! &=&\!\!\!\! x_t + \Delta g_1 (x^t, y^t, z^t) \\
y_{t+\Delta} \!\!\!\!&=&\!\!\!\! y_t + \Delta g_2 (x^t, y^t, z^t) \\
z_{t+\Delta} \!\!\!\!&=&\!\!\!\! z_t + \Delta g_3  (x^t, y^t, z^t)
\end{eqnarray*}
The causal dependencies can be depicted graphically (see Figure~\ref{fig:full_xyz}).  We can approximate this dynamical system by approximating the functions $\{g_1(x^t, y^t, z^t), g_2(x^t, y^t, z^t), g_3(x^t, y^t, z^t)\}$ and using fewer inputs.  For example, approximate $g_1 (x^t, y^t, z^t)$ with a function $g_1 ' (x^t)$.  One approximation for the system is:
\begin{eqnarray*}
x_{t+\Delta} \!\!\!\! &=&\!\!\!\! x_t + \Delta g_1 (x^t, y^t, z^t) \approx x_t + \Delta g_1 ' (x^t)  \\
y_{t+\Delta} \!\!\!\!&=&\!\!\!\! y_t + \Delta g_2 (x^t, y^t, z^t) \approx y_t + \Delta g_2 ' ( x^t, y^t)\\
z_{t+\Delta} \!\!\!\!&=&\!\!\!\! z_t + \Delta g_3  (x^t, y^t, z^t) \approx z_t + \Delta g_3 ' ( y^t, z^t),
\end{eqnarray*}
Figure~\ref{fig:tree_xyz} depicts the corresponding causal dependence
tree structure for these coupled differential equations.

A similar procedure can be used for stochastic processes, where the system is described in a time-evolving manner through conditional probabilities.  Consider three processes $\{\bX, \bY, \bZ\}$, formed by including i.i.d. noises $\{\epsilon_i,\epsilon_i',\epsilon_i''\}_{i=1}^n$   to the above dynamical system and relabeling the time indices (up to time $n$):
\begin{eqnarray*}
X_{i+1} \!\!\!\! &=&\!\!\!\! X_i + \Delta g_1 (X^i, Y^i, Z^i) + \epsilon_i \\
Y_{i+1} \!\!\!\!&=&\!\!\!\! Y_i + \Delta g_2 (X^i, Y^i, Z^i)  + \epsilon_i'\\
Z_{i+1} \!\!\!\!&=&\!\!\!\! Z_i + \Delta g_3  (X^i, Y^i, Z^i) + \epsilon_i''
\end{eqnarray*}
The system can alternatively be described through the joint distribution
\begin{eqnarray*}
&& \!\!\!\!\!\!\!\!\!\!P_{\bX, \bY, \bZ} (\bx, \by, \bz) = \\
&& \!\!\!\!\! \prod_{i=1}^n P_{X_i, Y_i, Z_i | X^{i-1}, Y^{i-1}, Z^{i-1}}(dx_i, dy_i, dz_i | x^{i-1}, y^{i-1}, z^{i-1}).
\end{eqnarray*}  Because of the causal structure of the dynamical system, given the full past, the present values are conditionally independent:
\begin{eqnarray*}
&& \!\!\!\!\!\!\!\!\!\!\!\!\!P_{\bX, \bY, \bZ} (d\bx, d\by, d\bz)= \\
&& \prod_{i=1}^n P_{X_i | X^{i-1}, Y^{i-1}, Z^{i-1}}(dx_i | x^{i-1}, y^{i-1}, z^{i-1}) \\
&& \qquad \times \ P_{Y_i | X^{i-1}, Y^{i-1}, Z^{i-1}}( dy_i | x^{i-1}, y^{i-1}, z^{i-1})  \\
&& \qquad \quad \times \ P_{Z_i | X^{i-1}, Y^{i-1}, Z^{i-1}}( dz_i | x^{i-1}, y^{i-1}, z^{i-1}). 
\end{eqnarray*}
Rewrite this using the notation of causal conditioning \eqref{eq:def:causal_cond} introduced by Kramer:
\begin{eqnarray*}
P_{\bX,\bY,\bZ}(d\bx,d\by,d\bz) \!\!&=&\!\! P_{\bX \parallel \bY\!, \bZ}(d\bx \parallel \by, \bz) P_{\bY \parallel \bX, \bZ}(d\by \parallel \bx, \bz) \\
&& \qquad \qquad \times P_{\bZ \parallel \bX, \bY}(d\bz \parallel \bx, \by).
\end{eqnarray*}  The dependence structure of this stochastic system is still represented by Figure~\ref{fig:full_xyz}.  We can apply a similar approximation to this system as before, corresponding to the structure of Figure~\ref{fig:tree_xyz}, with:
\begin{eqnarray*}
P_{\bX \parallel \bY, \bZ}(d\bx \parallel \by, \bz)  \!\!&\approx&\!\! P_{\bX}(d\bx) \\
P_{\bY \parallel \bX, \bZ}(d\by \parallel \bx, \bz) \!\!&\approx&\!\! P_{\bY \parallel \bX}(d\by \parallel \bx) \\
P_{\bZ \parallel \bX, \bY}(d\bz \parallel \bx, \by) \!\!&\approx&\!\! P_{\bZ \parallel \bY}(d\bz \parallel \by).
\end{eqnarray*}
Thus, our causal dependence tree approximation to these stochastic processes, denoted by $\PT$, is:
\[ P_{\allX}(d\allx) \!\approx \!\PT(d\allx) \triangleq P_{\bX}(d\bx) P_{\bY \parallel \bX}(d\by \!\parallel\! \bx) P_{\bZ \parallel \bY}(d\bz \!\parallel \!\by). \]
Note that, with this type of approximation, the processes are not mixed together  and, since the nodes represent processes, not individual variables, the graphical complexity remains low.
Another important characteristic is that the system we are approximating is causal and our approximation is causal, which might not have been the case if the Chow and Liu algorithm was applied.
We now consider the problem of finding the best causal dependence tree approximation using KL divergence as a measure of goodness.

\section{Main Result: Causal Dependence Tree Approximations}

\begin{figure}[t]
\centering
\includegraphics[width=.6\columnwidth]{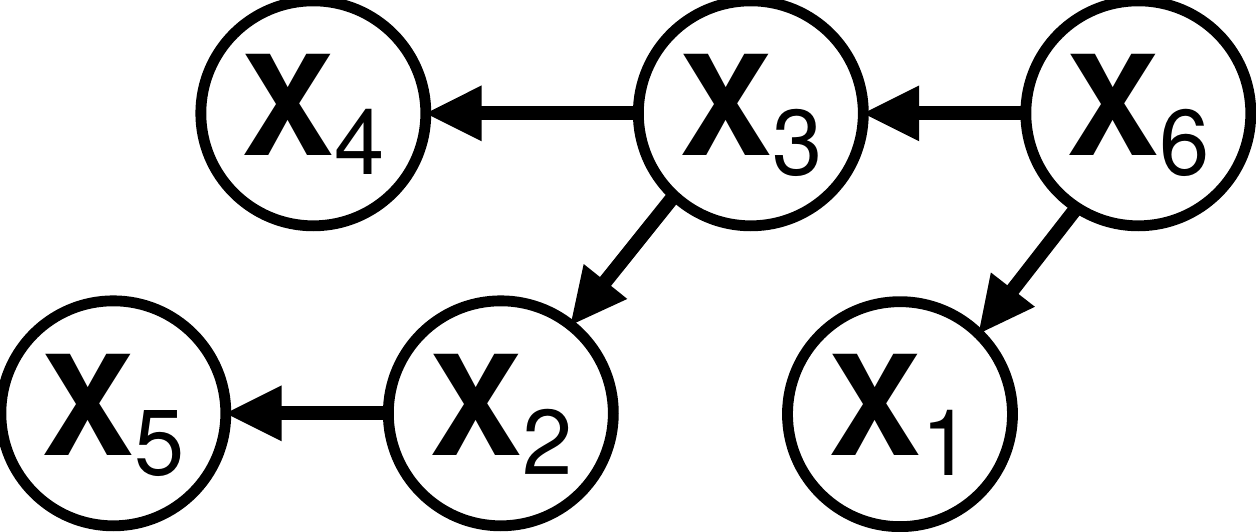}
\caption{Diagram of an approximating causal dependence tree structure.  In this example, \newline $\PT(d\allx) = P_{\bX_6}(d\bx_6) P_{\bX_1\parallel\bX_6}(d\bx_1 \parallel\bx_6) P_{\bX_3\parallel\bX_6}(d\bx_3 \parallel \bx_6)$\newline $\qquad \quad \times P_{\bX_4\parallel\bX_3}(d\bx_4 \parallel \bx_3) P_{\bX_2\parallel\bX_3}(d\bx_2 \parallel \bx_3) P_{\bX_5\parallel\bX_2}(d\bx_5 \parallel\bx_2)$. }
\label{fig:causal_dependence_tree}
\end{figure}

Consider the joint distribution $P_{\allX}$ of $m$ random processes $\{ \X_1, \ \X_2, \cdots,  \X_m$\}, each of length $n$.  For a given tree $T$ (defined by the functions $\pi(i)$ and $l(i)$ over the index set of the processes $i \in \{1, \cdots, m\}$), denote the corresponding causal dependence tree approximation as \begin{eqnarray}
\PT( d\allx ) \triangleq \prod_{i=1}^m P_{\bX_{\pi(i)}\parallel\bX_{\pi(l(i))}} (d\bx_{\pi(i)}\parallel \bx_{\pi(l(i))}). \label{eq:caus_dep_tree_appx}
\end{eqnarray}  An example of an approximating causal dependence tree, depicted as a directed tree, is shown in Figure~\ref{fig:causal_dependence_tree}.

As in Chow and Liu's work, KL divergence will be used to measure the ``goodness'' of the approximations.  Let $\PT(\allx)$ denote the causal dependence tree approximation of $P_{\allX}(\allx)$ for tree $T$. Let $\mcT$ denote the set of all causal dependence tree approximations for $P_{\allX}(\allx)$ and let $\tildeP(\allx)$ denote the product distribution 
\beqa
\tildeP (d\allx) &\triangleq&  \prod_{i=1}^m P_{\X_{i}}(d\bx_i),  \label{eqn:prod_dist} \\
                &=&\prod_{i=1}^m P_{\bX_{\pi(i)}} (d\bx_{\pi(i)}) \label{eqn:prod_dist_perm}
\eeqa 
which is equivalent to $P_{\allX}(\allx)$ when the processes are statistically independent.  Note that
\eqref{eqn:prod_dist_perm} holds for any permutation $\pi \in \Pi(m)$.

The following result for the causal dependence tree that minimizes the KL divergence holds:
\newtheorem{caus_dep_tree}[chow_liu]{Theorem} 
\begin{caus_dep_tree}
\label{caus_dep_tree}
\begin{eqnarray}
\!\!\!\!\!\!\!\!\! \argmin_{\PT \in \mcT}   \D ( P_{\allX} \parallel \PT ) \!\!\!\! &=& \!\!\!\! \argmax_{\PT \in \mcT}    \sum_{i=1}^m \I ( \X_{\pi(l(i))} \! \to \!  \X_{\pi(i)} ) \label{eq:caus_tree_result}
\end{eqnarray}
\end{caus_dep_tree}

\begin{proof}
Note that  $P_{\allX}$, $\PT$, $\tildeP$ all lie in $\probSimplex{\Omega}$, and moreover,
$P_{\allX} \ll \PT \ll \tildeP$. Thus, the Radon-–Nikodym  derivative  $\frac{dP_{\allX}}{d \tildeP}$ satisfies the chain rule \cite{royden1988real}: \[\frac{dP_{\allX}}{d \tildeP} = \frac{dP_{\allX}}{d \PT} \frac{d\PT}{d\tildeP}. \]  Taking the logarithm on both sides and rearranging terms results in:
\[ \log \frac{dP_{\allX}}{d \PT} = \log \frac{dP_{\allX}}{d \tildeP} - \log \frac{d\PT}{d\tildeP}.\]
\noindent Thus, \begin{eqnarray}
&& \!\!\!\!\!\!\!\!\!\!\!\!\!\!\!\!\!\!\!\! \argmin_{\PT \in \mcT}  \ \D ( P_{\allX} \parallel \PT ) \nonumber \\
&& \!\!\!\!\!\!\!\!\!\!\!\!\!\!\!\!\!\!\!\! = \argmin_{\PT \in \mcT}\ \E_{P_{\allX}} \left[  \log \frac{d P_{\allX}}{d \PT} \right] \nonumber \\
&& \!\!\!\!\!\!\!\!\!\!\!\!\!\!\!\!\!\!\!\!  = \argmin_{\PT \in \mcT}\ \E_{P_{\allX}} \left[   \log \frac{dP_{\allX}}{d \tildeP} \right] + \E_{P_{\allX}} \left[ - \log \frac{d\PT}{d\tildeP} \right] \nonumber \\
&& \!\!\!\!\!\!\!\!\!\!\!\!\!\!\!\!\!\!\!\!  = \argmax_{\PT \in \mcT} \ \E_{P_{\allX}} \left[  \log \frac{d \PT}{d\tildeP} \right] \label{eq:main_res:pf1} \\
&& \!\!\!\!\!\!\!\!\!\!\!\!\!\!\!\!\!\!\!\! = \argmax_{\PT \in \mcT} \ \sum_{i=1}^m \int_{\allx} \log \frac{dP_{\bX_{\pi(i)}\parallel\bX_{\pi(l(i))}=\bx_{\pi(l(i))}}}{dP_{\bX_{\pi(i)}}}
P_{\allX}(d\allx) \label{eq:main_res:pf2} \\
&& \!\!\!\!\!\!\!\!\!\!\!\!\!\!\!\!\!\!\!\! = \!\!\argmax_{\PT \in \mcT} \!\!\sum_{i=1}^m \!\!\int_{\bx} \!\!\!\!\kldist{P_{\bX_{\pi(i)}\parallel\bX_{\pi(l(i))}=\bx}}{P_{\bX_{\pi(i)}}} \!\!P_{\bX_{\pi(l(i))}}(d\bx)
\label{eq:main_res:pf2a} \\
&& \!\!\!\!\!\!\!\!\!\!\!\!\!\!\!\!\!\!\!\!  = \argmax_{\PT \in \mcT} \   \sum_{i=1}^{m} \I ( \X_{\pi(l(i))} \to  \X_{\pi(i)} ),  \label{eq:main_res:pf3}
\end{eqnarray} where \eqref{eq:main_res:pf1} follows from $\frac{dP_{\allX}}{d\tildeP}$ not depending on $\PT$;
 \eqref{eq:main_res:pf2} follows from \eqref{eq:caus_dep_tree_appx} and \eqref{eqn:prod_dist_perm}; \eqref{eq:main_res:pf2a} follows from \eqref{eqn:defn:divergence};
 and \eqref{eq:main_res:pf3} follows from \eqref{eqn:defn:DirectedInformation}.

\end{proof}
Thus, finding the optimal causal dependence tree in terms of KL distance is equivalent to maximizing a sum of directed informations.  Also note that when $n=1$, there is an equivalence between this and Chow and Liu's result:
\newtheorem{ours_CL_equiv}[chow_liu]{Corollary} 
\begin{ours_CL_equiv}
When $n=1$, Theorem~\ref{caus_dep_tree} reduces to Theorem~\ref{chow_liu}.
\label{ours_CL_equiv}
\end{ours_CL_equiv}

Similar to Chow and Liu's result, only the pairwise interactions between the processes need to be known or estimated to identify the best approximation for the whole joint.  Two estimators for directed information from one process to another have recently been proposed.  A parametric approach based on the law of large numbers for Markov chains and minimum description length is presented in \cite{quinn2010estimating}.  A universal estimation approach based on context weighting trees is presented in \cite{zhao2010universal}.

\section{A Low Complexity Algorithm for Finding the Optimal Causal Dependence Tree}

In Chow and Liu's work, Kruskal's minimum spanning tree algorithm  performs the  optimization procedure efficiently, after having computed the mutual information between each pair of variables \cite{chow1968approximating}.  A similar procedure can be done in this setting.  First, compute the directed information between each ordered pair of processes.  This can be represented as a graph, where each of the nodes represents a process.  This graph will have a directed edge from each node to every other node (thus is a complete, directed graph), and the value of edge from node $\X$ to node $\Y$ will be $\I(\X \to \Y)$.

There are several efficient algorithms which can be used to find the maximum weight (sum of directed informations) directed tree of a directed graph \cite{gabow1986efficient}, such as Chu and Liu \cite{chu1965shortest} (which was independently discovered by Edmonds \cite{edmonds1967optimum} and Bock \cite{bock1971algorithm}) and a distributed algorithm by Humblet \cite{humblet1983distributed}.  Note that in some implementations, a root is required a priori.  For those, the implementation would need to be applied for each node in the graph as a root, and then the directed tree which has maximal weight among all of those would be selected.  Chu and Liu's algorithm has runtime of $\mathcal{O}(m^2)$ \cite{gabow1986efficient}.  The total runtime of this procedure is $\mathcal{O}(m^3)$.

\section{Properties of causal dependence trees}

Now we will consider some of the differences between Chow and Liu dependence trees and causal dependence trees in terms of variable dependencies and storage requirements.

\subsection{Dependencies between variables}

\begin{figure}[t]
\centering
\mbox{\!\!
\subfigure[Variable dependence structure for a full joint distribution.] {\label{fig:vars_graph:full}\includegraphics[width=.33\columnwidth]{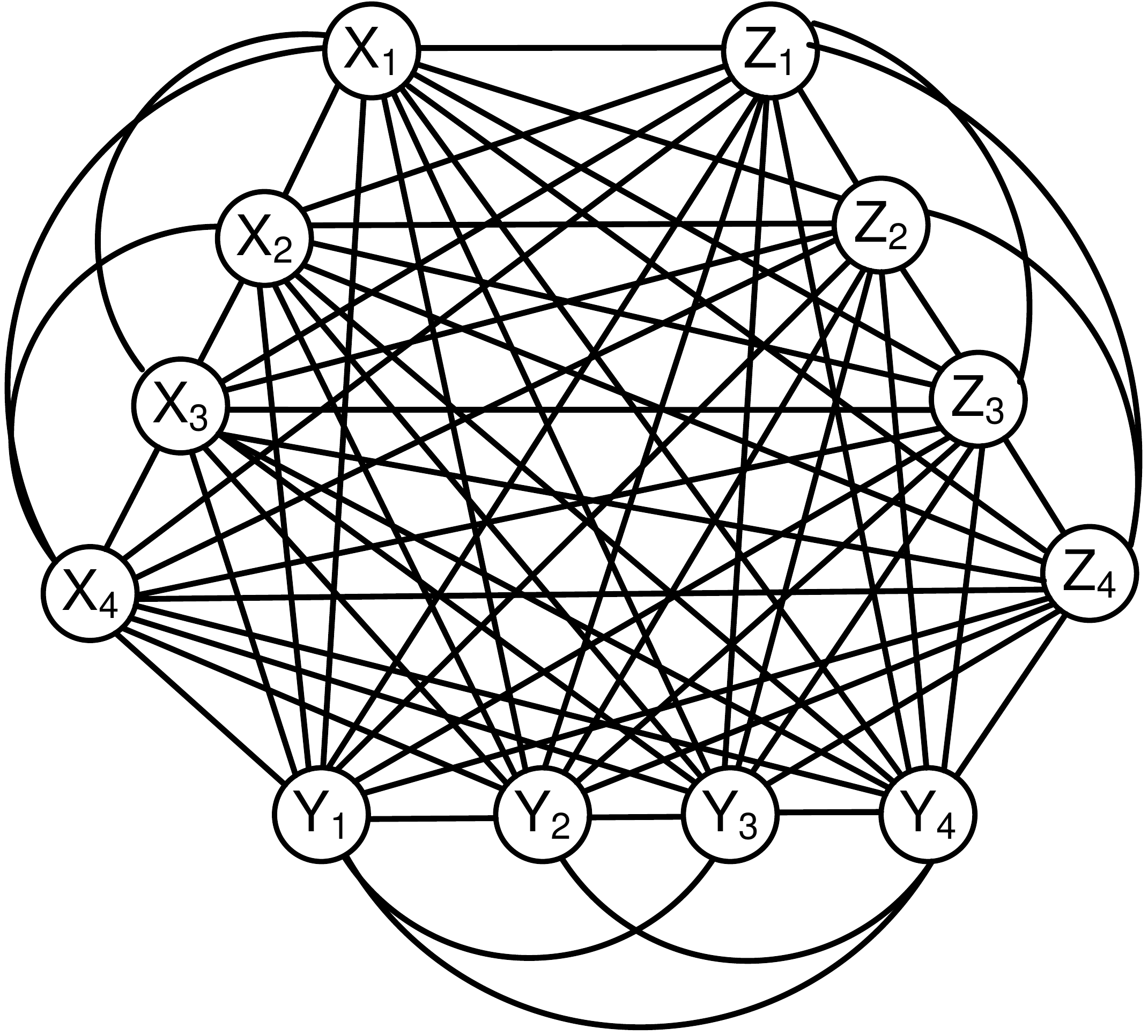}}
\ \subfigure[Variable dependence structure for a particular Chow and Liu dependence tree approximation.] {\label{fig:vars_graph:CLtree}\includegraphics[width=.33\columnwidth]{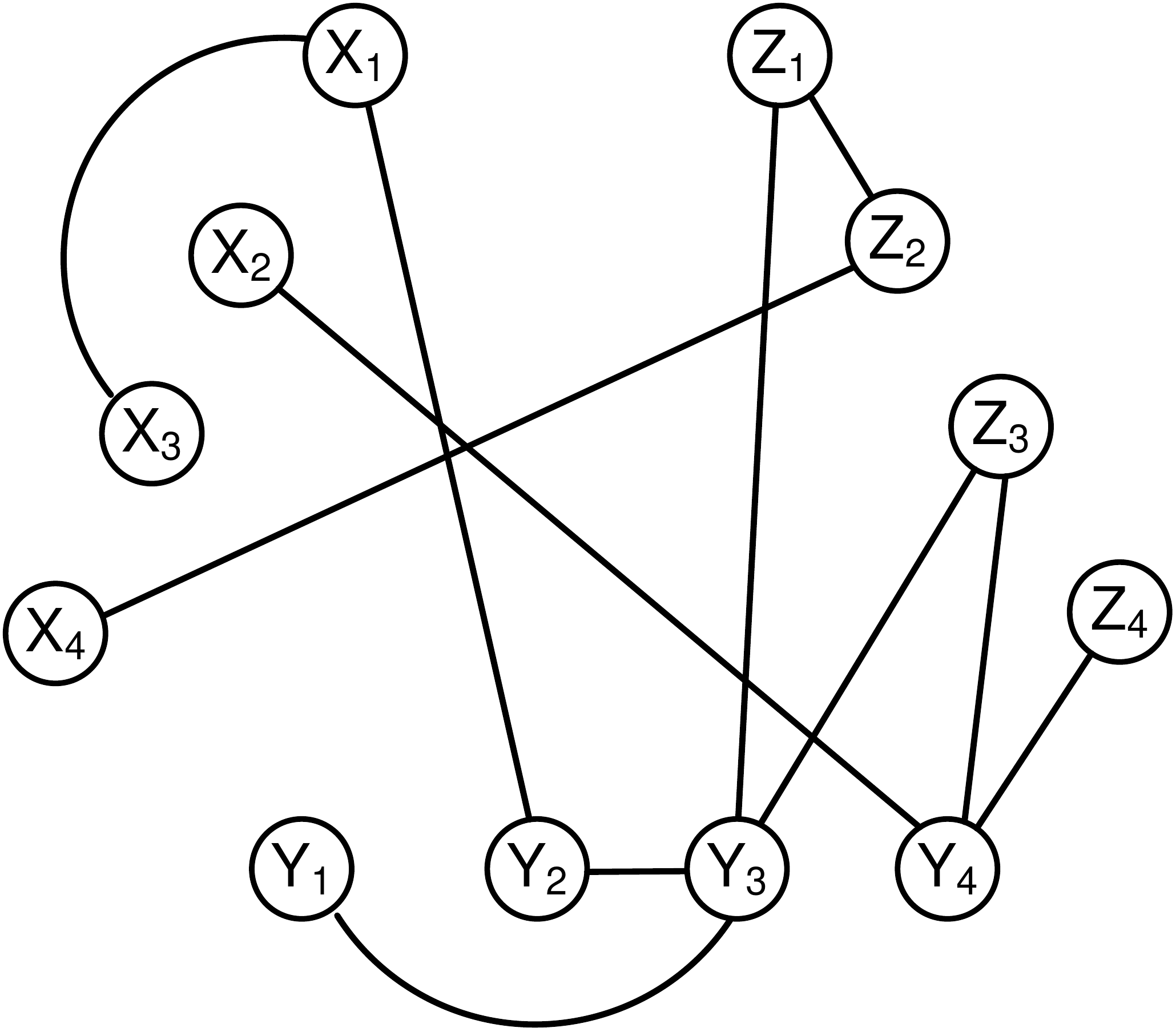}}
\subfigure[Variable dependence structure for a causal dependence tree approximation
$\widehat{P}_{\bX,\bY,\bZ}(d\bx,d\by,d\bz) =$ \newline $ P_{\bX}(d\bx) P_{\bY \parallel \bX}(d\by \parallel \bx)$ \newline $P_{\bZ \parallel \bY}(d\bz \parallel \by).$]{\label{fig:vars_graph:tree}\includegraphics[width=.33\columnwidth]{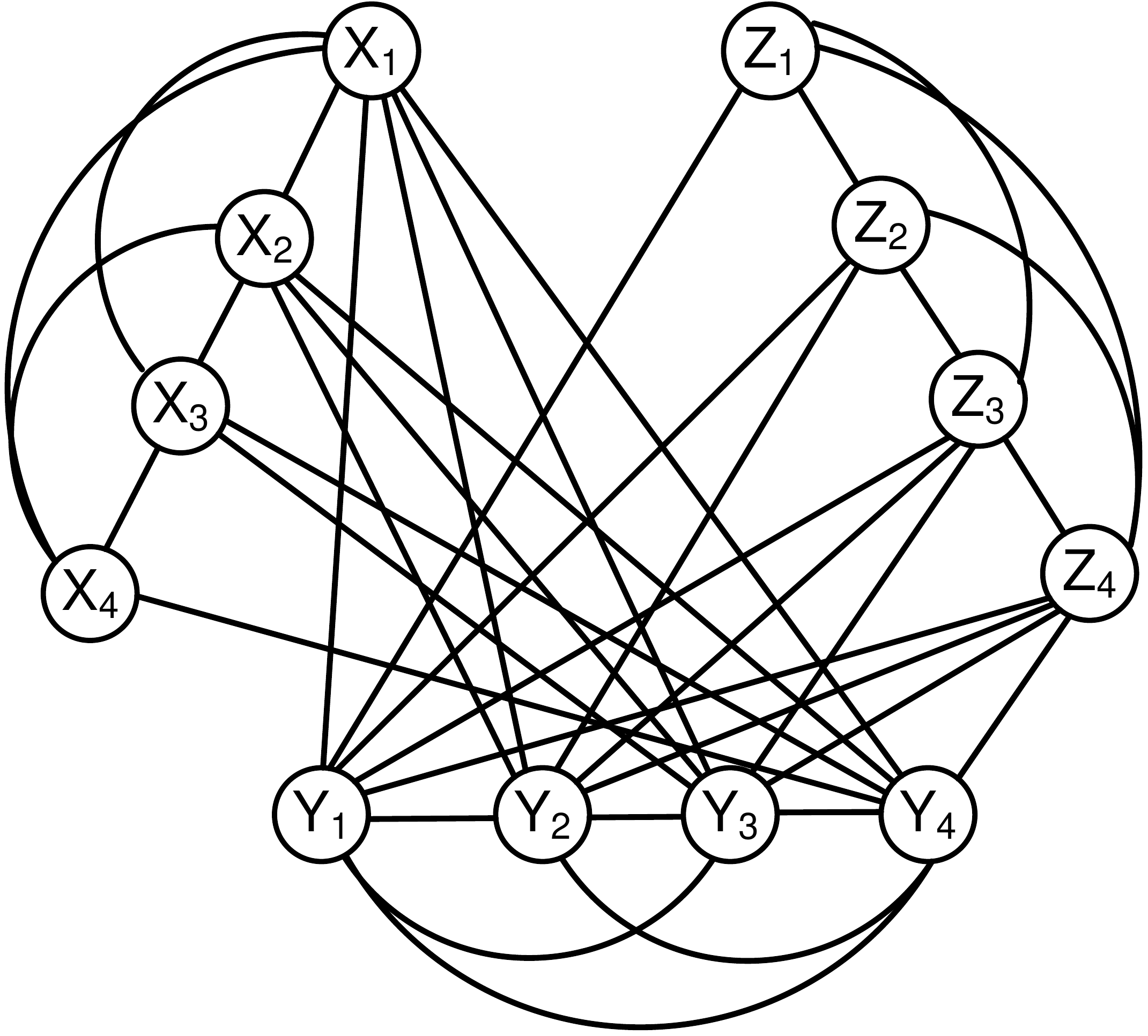}}
}
\caption{The variable dependence structures for a full joint distribution, a Chow and Liu dependence tree approximation, and a causal dependence tree approximation, for a set of three random processes with four timesteps.}
\label{fig:vars_graph:comp}
\end{figure}

Causal dependence trees have a simple graphical representation for random processes, unlike Chow and Liu dependence trees.  For causal dependence trees, the \emph{processes} are represented by nodes, not the variables.  However, the dependencies between variables induced by causal dependence trees can also be graphically represented.  An example showing dependencies between variables for the full joint distribution, a Chow and Liu dependence tree approximation, and a causal dependence tree approximation, in a set of three random processes with four timesteps, is depicted in Figure~\ref{fig:vars_graph:comp}.

The graph of the variable dependence structure induced by a causal dependence tree approximation (Figure~\ref{fig:vars_graph:tree}) is not necessarily a tree.  It is a structured subgraph of the variable dependence structure of the full joint (Figure~\ref{fig:vars_graph:full}).  In particular, a variable is allowed to have dependencies with all of the previous variables in its process and those in the past of the process being causally conditioned on.  Consequently, the induced subgraph of the variables from a single process, such as $\{Y_1, Y_2, Y_3, Y_4\}$ form a complete graph.  In general, the set of possible Chow and Liu dependence trees (any tree on the variables) does not intersect with the set of possible causal dependence trees.  In the limiting case of $n=1$, the sets of possible trees are the same (see Corollary~\ref{ours_CL_equiv}).

Even though the graph of variable dependencies for a causal dependence tree is more complex than that of a Chow and Liu dependence tree, it is significantly less complex then a full joint distribution.  Consider a network of $m$ random processes over $n$ timesteps.  There are $mn$ variables total.  For the graph of dependencies between variables for the full joint distribution, there are $\mathcal{O}({mn \choose 2}) = \mathcal{O}(m^2 n^2)$ edges.  The Chow and Liu dependence tree has $mn-1$ or $\mathcal{O}(mn)$ edges.  The graph of dependencies between variables for a causal dependence tree distribution has a complete graph for each process ($m {n \choose 2}$ edges), as well as $k$ edges between a variable with index $k$ to the current and all of the previous $k-1$ variables in the process being causally conditioned on.  Since there are $m-1$ processes which are causally conditioned on one other, there are \[(m-1) \sum_{k=1}^{n} k = (m-1) \frac{n(n+1)}{2}\] edges between variables of different processes.  Consequently, the causal dependence tree has \[\mathcal{O} \left( m {n \choose 2} + (m-1) \frac{n(n+1)}{2} \right) = \mathcal{O}(mn^2)\] edges total.  These extra dependencies (edges) allow causal dependence trees to incorporate more dynamics of the system that pertain to how the processes evolve depending on their own past and possibly the past of other processes.

\subsection{Storage requirements}
One of the significant aspects of using the original Chow and Liu algorithm is the reduction in storage needed for the approximation.  We will now examine the reduction in storage for causal dependence trees.  Let $m$ denote the number of processes, and $n$ the length in time.  There are $mn$ variables total.  For simplicity, assume each variable is over a finite alphabet of size $|\calX| < \infty$.  The full distribution requires $\mathcal{O} (|\calX|^{mn})$ storage, since there are $|\calX|^{mn}$ realizations, each with a possibly unique probability.

The Chow and Liu algorithm approximates the full joint with a product of second order distributions \cite{chow1968approximating}.  For example, given a joint distribution on six random variables, $P_{X^6}(dx^6)$, the Chow and Liu algorithm might approximate it as in Figure~\ref{fig:dependence_tree} with the following:
\begin{eqnarray*}
\widehat{P}_{X^6}(dx^6) \!\!\!\! &=& \!\!\!\! P_{X_6}(dx_6) P_{X_1|X_6} (dx_1|x_6) P_{X_3|X_6}(dx_3|x_6)  \\
&& \!\!\!\!\!\!\!\!\!\!\!\!\!\!\!\!\!\!\!\!\!\!\!\!\times \ P_{X_4|X_3}(dx_4|x_3)P_{X_2|X_3}(dx_2|x_3) P_{X_5|X_2}(dx_5|x_2),
\end{eqnarray*}
or another product of this form.  Each second order distribution requires $\mathcal{O}(|\calX|^2)$ storage, and there are $mn-1$ of them, one for each variable except the first, which has first order distribution.  Thus, the total storage required for a Chow and Liu dependence tree approximation is $\mathcal{O} (mn |\calX|^{2})$, which is linear in both the number of processes and time.

The causal dependence tree approximation has a much simpler graphical representation than the Chow and Liu procedure in the context of random processes.  However, it largely does not restrict dependencies within each process and between processes where causal dependencies are kept.  For example, consider three processes $\{\bX, \bY, \bZ\}$ with a causal tree approximation \[\widehat{P}_{\bX,\bY,\bZ}(d\bx,d\by,d\bz)  =P_{\bX}(d\bx)P_{\bY \parallel \bX}(d\by \parallel \bx) P_{\bZ \parallel \bY} (d\bz \parallel \by).\]  This can be expanded into a product of conditional probabilities with increasing time
\begin{eqnarray*}
&& \!\!\!\!\!\!\!\!\!\!\!\!\!\!\!\!\widehat{P}_{\X,\Y,\Z}(d\bx,d\by,d\bz) = \\
&& \quad  \prod_{i=1}^n P_{X_i |X^{i-1}}(dx_i | x^{i-1}) P_{Y_i | Y^{i-1}, X^{i}}(dy_i | y^{i-1}, x^{i}) \nonumber \\
&& \qquad  \quad \times P_{Z_i | Z^{i-1}, Y^{i}}(dz_i | z^{i-1}, y^{i}) \nonumber\end{eqnarray*}

The final terms have many dependencies.  A variable is allowed to depend on the full past of its own process and the process that it is causally conditioned upon.  The storage for the whole causal tree approximation will be dominated by the storage required for these terms.  For each of these $m-1$ final terms (conditioned on full past of two processes), $\mathcal{O}(|\calX|^{2n})$ storage is required, so the total storage necessary is $\mathcal{O} (m |\calX|^{2n})$.  Thus, the storage for causal dependence trees is exponentially worse than that for Chow and Liu dependence trees, but exponentially better than storing the full joint distribution.

\section{Example}


Let us illustrate the proposed algorithm with a binary hypothesis testing example.  We construct two networks of jointly gaussian random processes according to a generative model.  Next, we apply the above procedure to form causal dependence tree approximations for both networks.  Additionally, we apply the original Chow and Liu procedure to develop dependence tree approximations.  Subsequently, the data generated from the original distributions is used in binary hypothesis testing (using log likelihood ratios with a threshold parameter).  The performance of the causal dependence tree approximations in binary hypothesis testing is compared to that of the original distributions and that of the Chow and Liu dependence trees.

The formula to compute directed information from a random process $\X$ to a random process $\Y$, where $\X$ and $\Y$ are jointly gaussian random processes each of length $n$, is:
\begin{eqnarray*}
I(\X \to \Y) &=& \sum_{i=1}^n I(Y_i;X^{i} | Y^{i-1}) \nonumber \\
&=& \frac{1}{2} \log\left[  |K_{Y^n}| \right] - \sum_{i=1}^n \frac{1}{2} \log\left[ \frac{|K_{Y^i, X^{i}}|}{|K_{Y^{i-1}, X^{i}}|} \nonumber \right], 
 \end{eqnarray*} where $|K_{Y^i, X^{i}}|$ is the determinant of the covariance matrix for the variables $\{Y^i, X^{i}\}$.  The last line follows from \cite{cover2006elements}. We will now construct the networks, and then use this formula to calculate causal dependence tree approximations for the networks.

\begin{figure}[t]
\centering
\mbox{\subfigure[Causal dependency graph for first generative model ($H_0$).]{\label{fig:bin_hyp_model_H0}\includegraphics[width=.35\columnwidth]{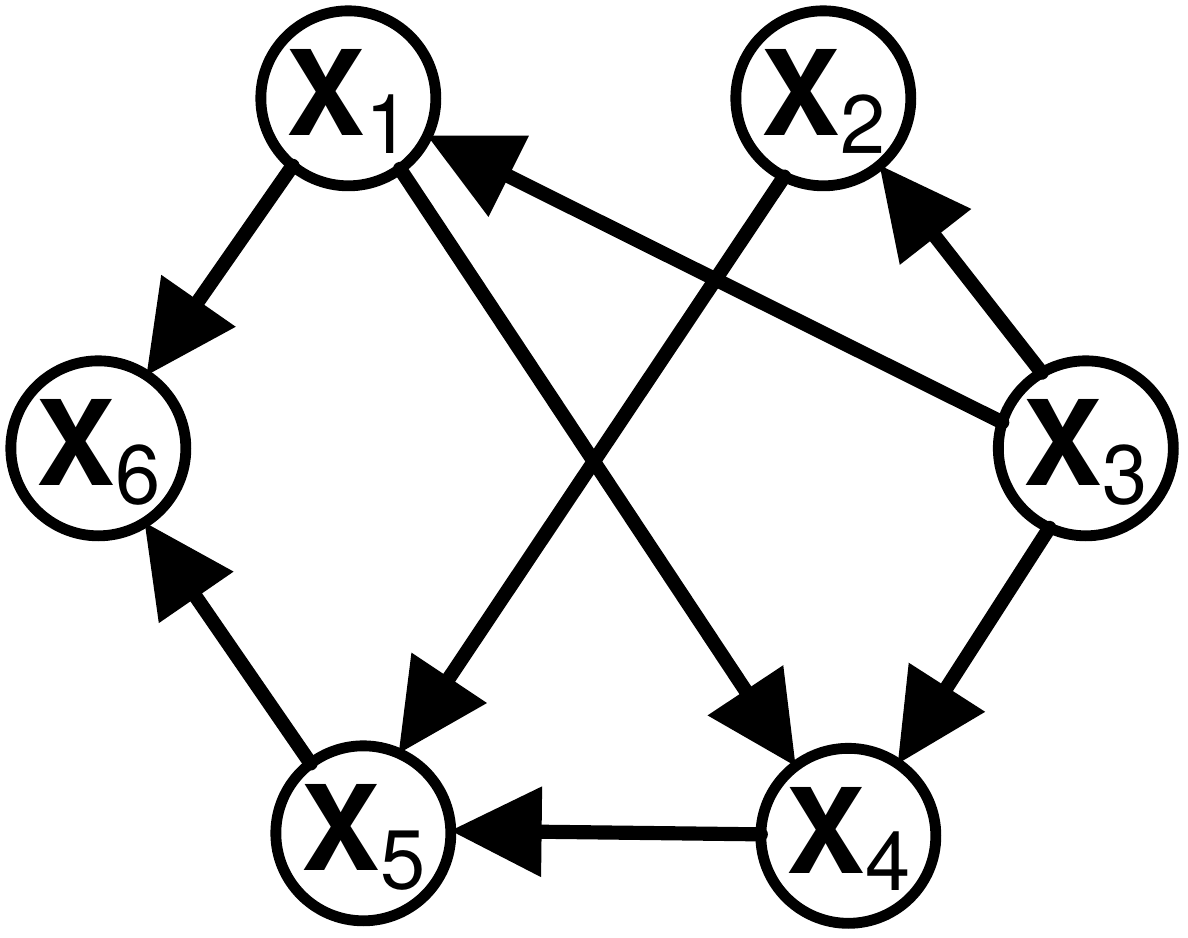}}\qquad\qquad
\subfigure[Causal dependency graph for second generative model ($H_1$).]{\label{fig:bin_hyp_model_H1}\includegraphics[width=.35\columnwidth]{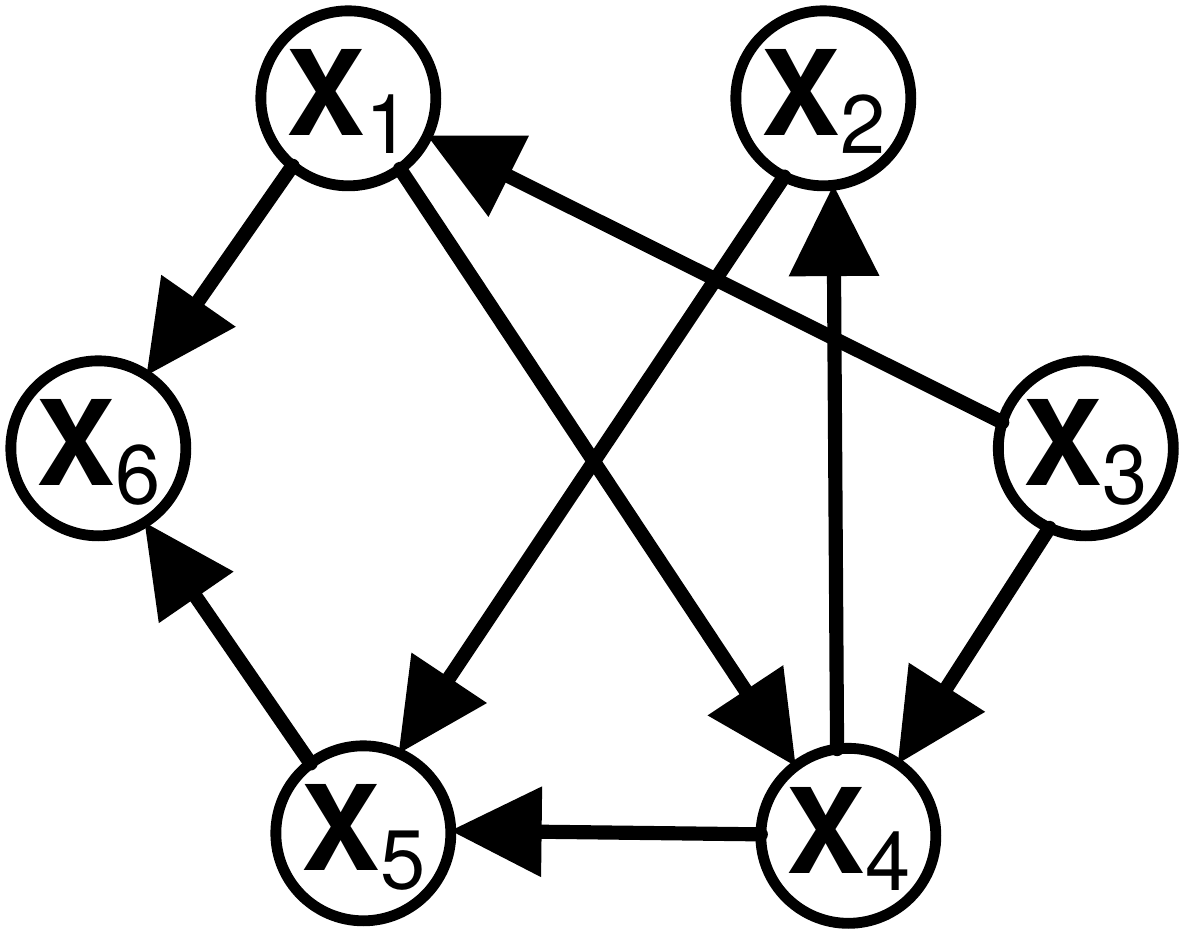}}}
\caption{Graphs of the causal dependencies between the processes in the full joint distributions for the two generative models.  The dependence structures are topologically similar.}
\label{fig:bin_hyp_models}
\end{figure}

\begin{figure}[t]
\centering
\mbox{\subfigure[Causal dependence tree approximation for the first generative model ($H_0$).]{\label{fig:bin_hyp_model_tree_H0}\includegraphics[width=.35\columnwidth]{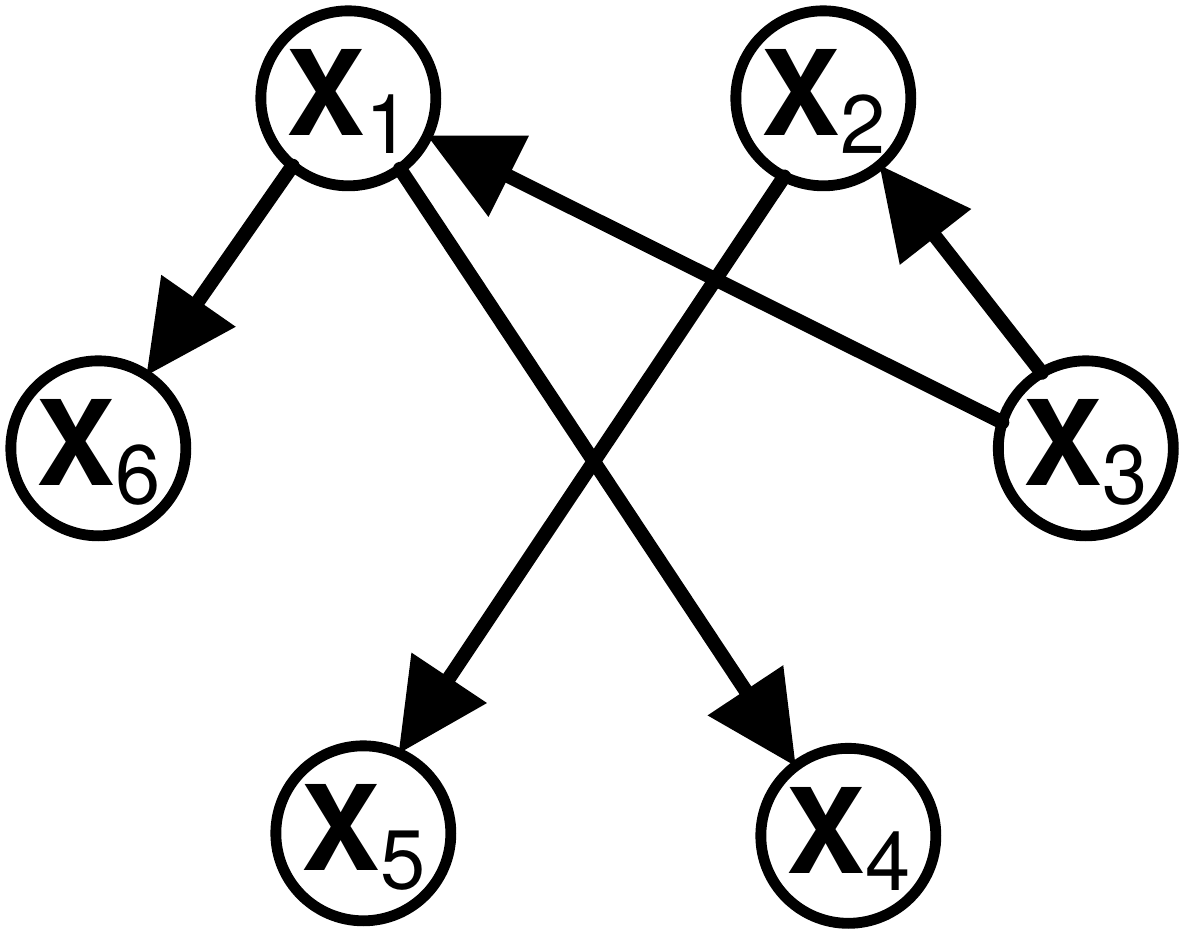}}\qquad\qquad
\subfigure[Causal dependence tree approximation for the second generative model ($H_1$).]{\label{fig:bin_hyp_model_tree_H1}\includegraphics[width=.35\columnwidth]{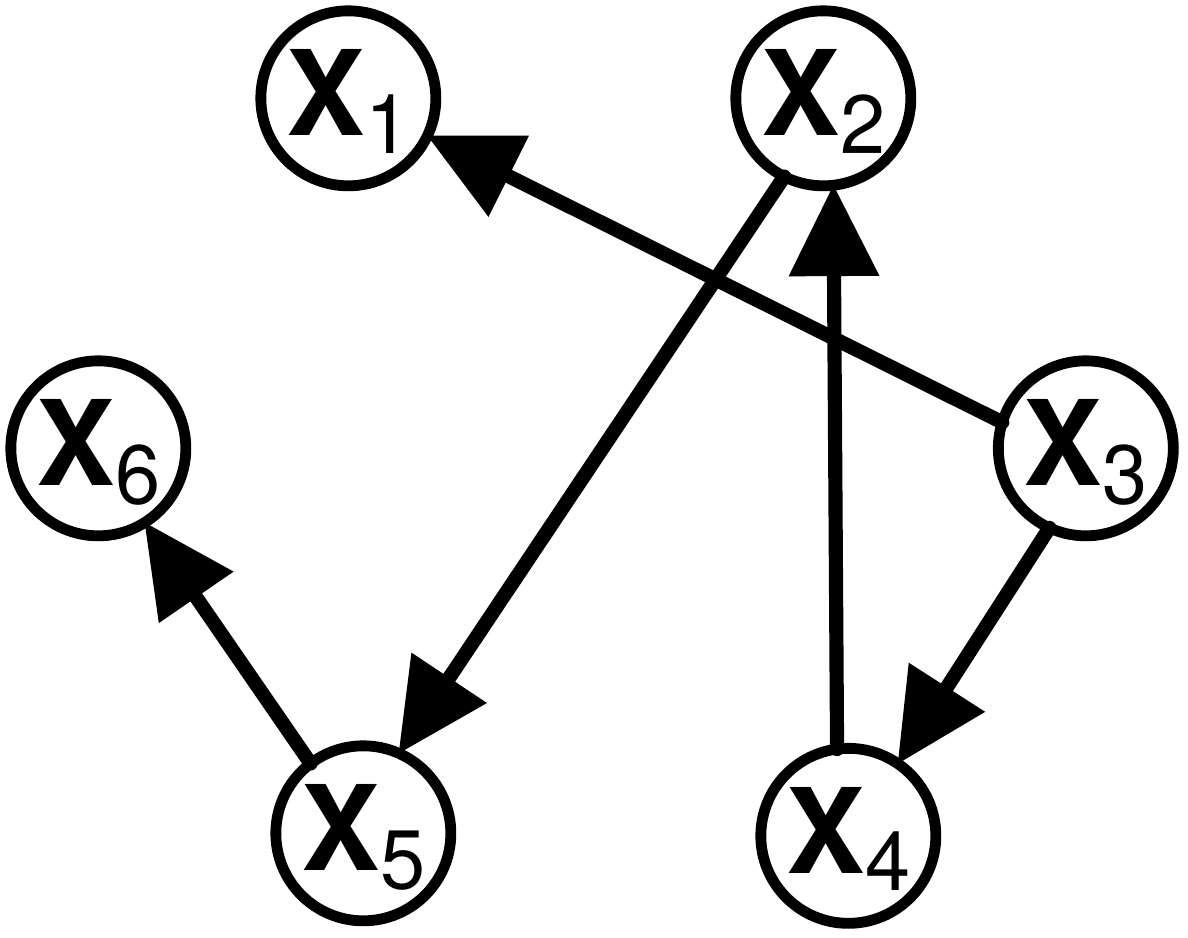}}}
\caption{Graphs of the causal dependence tree approximations for both of the generative model networks.  Despite the topological similarities of the dependence graphs for the original distributions, these approximations are topologically distinct.}
\label{fig:bin_hyp_models_trees}
\end{figure}
 
Let $\X^6$ denote six jointly gaussian, zero mean random processes.  We specified two generative models, where each process at time $i$ was a linear combination of a subset of the recent past of the other processes plus independent gaussian noise.  Letting $\overline{\X}$ denote a column vector containing all the variables, and $\N$ a column vector of independent normal noise, we specified the matrix $\A$ in: \[\overline{\X} = \A \overline{\X} + \N. \]  To obtain the full covariance matrix for $\X$, isolate $\overline{\X}$: \[  \overline{\X} = (\bfI - \A)^{-1} \N\] and compute $\overline{\X} \  \overline{\X}^{T}$.  Data can be generated for $\overline{\X}$ by first generating $\N$ and then linearly transforming the result.  The generative model graphs (with directed arrows depicting the causal dependencies) are shown in Figures \ref{fig:bin_hyp_model_H0} and \ref{fig:bin_hyp_model_H1}.  We applied the procedure to these two networks of jointly gaussian random processes, and the resulting causal dependence tree structures are depicted in Figures \ref{fig:bin_hyp_model_tree_H0} and \ref{fig:bin_hyp_model_tree_H1}.  We also used the Chow and Liu procedure to develop dependence tree approximations.  To compute the Chow and Liu dependence tree approximation, we used publicly available code \cite{li2009maximum}.  The number of dependencies between variables for the full joint distribution, the causal dependence tree approximations, and the Chow and Liu dependence tree approximations were $1770$, $495$, and $59$ respectively.

\begin{figure}[t]
\centering
\includegraphics[width=\columnwidth]{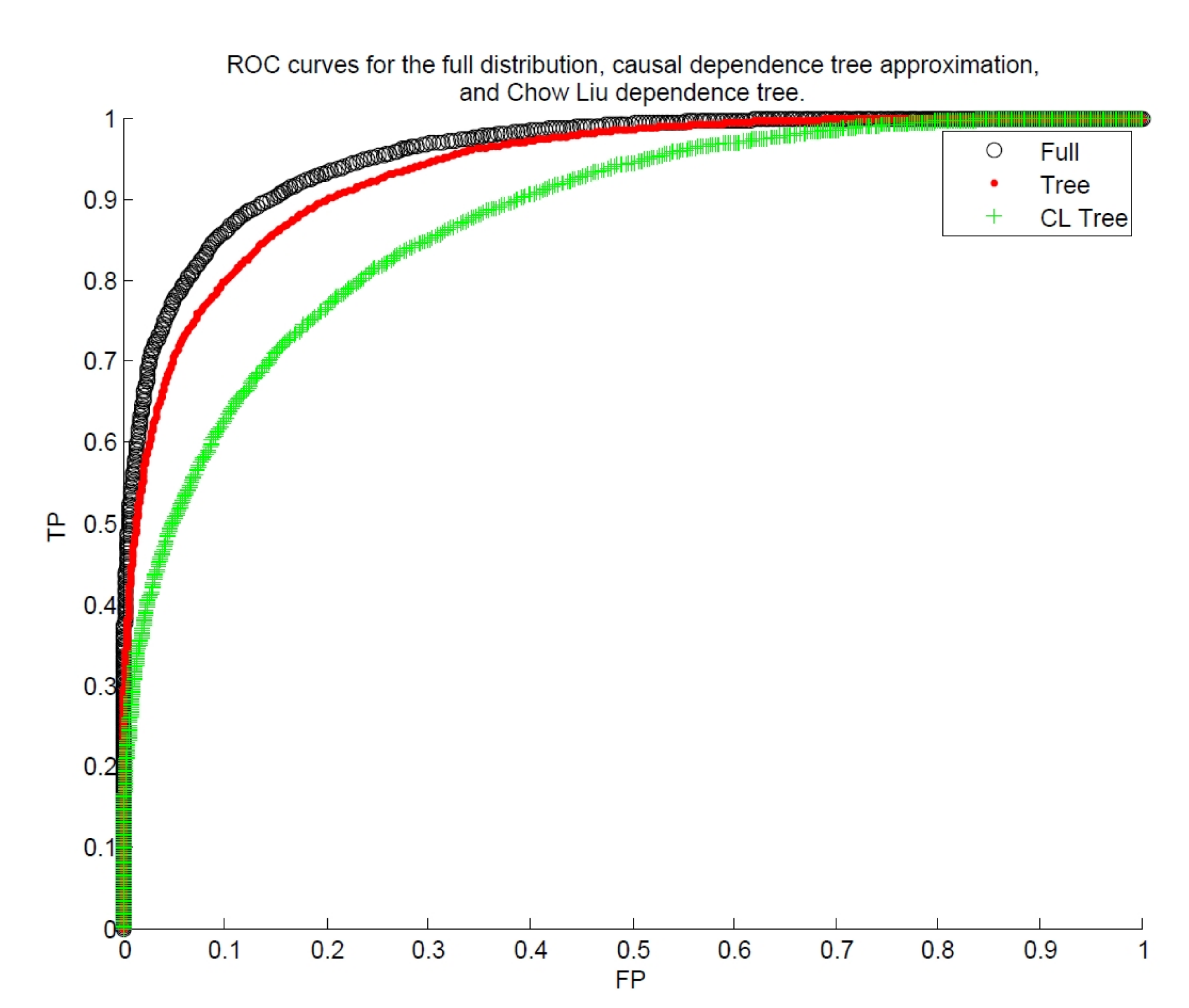}
\caption{ROC curves for the full distributions (top-most), the causal dependence tree approximations (middle), and the Chow and Liu dependence tree approximations (bottom-most) in binary hypothesis testing.}
\label{fig:bin_hyp_ROC}
\end{figure}

Next, we generated data $10 \ \!\!000$ times from both original distributions and performed binary hypothesis testing (using log likelihood ratios with a threshold $\tau$) with the original distributions, the causal dependence tree approximations, and the Chow and Liu dependence tree approximations.  Figure~\ref{fig:bin_hyp_ROC} depicts the corresponding ROC curves.  The causal dependence tree approximations, despite the significant reduction in structure, still perform well in this task.  Also, their performance is significantly better than that of the Chow and Liu dependence tree approximations.  

\section*{Acknowledgements}
The authors thank Mavis Rodrigues for her assistance with computer simulations.

\bibliographystyle{IEEEtran}

\end{document}